\pdfoutput=1
\documentclass[twoside,10.5pt]{article}
\usepackage[
backend=bibtex,
style=nature,
sorting=none
]{biblatex}
\usepackage{mathrsfs}
\usepackage{pifont}
\usepackage{amsmath}
\usepackage{amsthm}
\usepackage{txfonts}
\usepackage{geometry}
\usepackage{latexsym}
\usepackage{amssymb}
\usepackage{graphicx}
\graphicspath{ {figures/} }
\usepackage{wrapfig}
\usepackage[justification=centering]{caption}
\usepackage[font={small,it}]{caption}
\usepackage{verbatim}
\usepackage{geometry}
\usepackage{xcolor} 
\begin{document}
\begin{center}
{\textbf{Why Mutant Allele Frequencies in Oncogenes Peak Around 0.40 and Rapidly Decrease?}}\\[20pt] 
{\textit{Kasthuri Kannan\footnote[1]{New York University - School of Medicine}$^{,}$\footnote[2]{Corresponding author} and Adriana Heguy\footnotemark}}\\[20pt]
\end{center}

\textbf{Abstract}

The mutant allele frequencies in oncogenes peak around $0.40$ and rapidly decrease. In this article, we explain why this is the case. Invoking a key result from mathematical analysis in our model, namely, the inverse function theorem, we estimate the selection pressures of the mutations as a function of germline allele frequencies. Under complete dominance of oncogenic mutations, this selection function is expected to be linearly correlated with the distribution of the mutant alleles. We demonstrate that this is the case by investigating the allele frequencies of mutations in oncogenes across various cancer types, validating our model for mean effective selection. Consistent with the population genetics model of fitness, the selection function fits a gamma distribution curve that accurately describes the trend of the mutant allele frequencies. While existing equations for selection explain evolution at low allele frequencies, our equations are general formulas for natural selection under complete dominance operating at all frequencies. We show that selection exhibits linear behavior at all times, favoring dominant alleles with respect to the change in recessive allele frequency. Also, these equations show, selection behaves like power-law against the recessive alleles at low dominant allele frequency. 



\newpage

\begin{center}
{\textbf{Why Mutant Allele Frequencies in Oncogenes Peak Around 0.40 and Rapidly Decrease?}}\\[20pt] 
{\textit{Kasthuri Kannan\footnote[1]{New York University - School of Medicine}$^{,}$\footnote[2]{Corresponding author} and Adriana Heguy\footnotemark}}\\[20pt]
\end{center}

\textbf{1. Introduction}

Cancer is an evolutionary disease with mutation acting as a unit of selection and, therefore, quantifying selection is necessary. Since the distribution of the allele frequencies of somatic mutations reflects the nature of selection underlying the mutant clones, modeling this distribution is essential. Recently, Williams et al. \cite{williams} showed neutral tumor evolution results in a power-law distribution of the mutant allele frequencies, and this law fits $303$ of $904$ cancers of various types. While neutral evolution remains an important aspect in several cancer types, the distribution of the allele frequencies of mutations in genes that undergo positive selection, has not been determined so far. This is difficult to model because the allele frequencies of positively selected mutations are related to their functional consequences and therefore we have to take into account the degree of dominance exhibited by these mutations. However, in the case of oncogenes, which are primarily dominant, determining the distribution of the mutant allele frequencies should be feasible. It is worth noting that in the context of human polymorphisms, the allele frequencies of new deleterious mutations has been studied and fitness distribution is shown to follow a gamma distribution \cite{eyre}. Nonetheless, the fitness function and distribution of allele frequencies of oncogenic mutations in tumors have not been described. 

The mutant allele frequencies in oncogenes peak around $0.40$ and rapidly decrease. We built a mathematical model to describe the trend of these frequencies. We assumed complete dominance of oncogenic mutations, although we realize dominance (or partial dominance) can be modeled as a function of the functional impact of these mutations, which can be derived from algorithms such as PolyPhen and SIFT \cite{polyphen} \cite{sift}. The scope of this article is restricted to describe the general tendency of the frequencies rather than considering the impact of the mutations on their frequencies. By taking advantage of a key result in mathematical analysis, namely, the inverse function theorem, we estimate the mean effective selection of the mutations as a function of germline allele frequencies. Under complete dominance of oncogenic mutations, this selection function is expected to be linearly correlated with the distribution of the mutant alleles. We demonstrate that this is the case by investigating the allele frequencies of mutations in oncogenes across various cancer types, validating our model for mean effective selection. Consistent with population genetics model of fitness, the selection function fits a gamma distribution curve that accurately describes the trend of the mutant allele frequencies. 

This model infers mean effective selection for oncogenic mutations without considering other alterations in the DNA that could change the dynamics of the tumor micro-environment. Moreover, this measure is an \textit{effective selection coefficient} in the sense that it is selection coefficient with respect to gain in mutant allele frequencies. Combining this estimate with other modifications, such as copy number changes, and integrating the functional impact of the resulting protein will help in understanding the evolution of the mutant clones in various tumors. Although the equations that we derive are currently applied in the context of oncogenes, these are general formulas for natural selection under complete dominance operating at all frequencies. While being consistent with known formulas that explain evolution at low frequencies, one of these equations show that under low frequencies of the dominant alleles, selection against recessive alleles behave in a power-law like manner, reiterating the powerful role of natural selection. This would also explain the reason some tumors undergo rapid clonal evolution. Further, at high frequencies of the dominant alleles, the linear expansion exhibited by selection with respect to change in recessive germline allele frequencies could partly be the reason behind drug resistance, under dominance.

\textbf{2. Determining Selection Coefficients}

\noindent We use the standard model described in \cite{falconer} for selection under complete dominance. An illustration of this model for mutations when the frequencies are not time dependent is shown in Supplementary Figure 1. Under this model, if $s$ is the coefficient of selection, $p$ and $q$ are the allele frequencies with $p+q=1$, the new genotype frequencies $f$, in terms of $p$ and $q$ are given by,

\begin{equation}
    \label{eq:basic_mut_frequency}
    f(p)=\frac{p^2+pq}{1-sq^2}; \hspace{3mm} f(q)=1-f(p)=\frac{q-sq^2}{1-sq^2}
\end{equation}

Using this model for time-dependent allele frequencies, we derive a new equation for the number of somatic mutations in terms of the germline allele frequency.

\noindent At time $t$, let $M$ be a mutation with $p_{t}^2,2p_{t}q_{t}$ and $q_{t}^2$ as allele frequencies of mutant homozygous, mutant heterozygous and germline genotypes of $M$, respectively, with $p_{t}+q_{t}=1$. Let the strength of the selection be expressed as a coefficient of selection, $s$, which is proportional to the reduction of the germline genotype, compared to the mutant genotypes which are favored for the tumor growth. In reality, the selection coefficient should be a function of time. However, we can consider $s$ as mean selection acting over time and so the time dependence can be omitted. Thus, in our model, we assume the frequencies are time-dependent, but the selection is time-independent. If the fitness of the homozygous and heterozygous mutant genotypes are taken to be $1$ as it is likely the case in oncogenes, the fitness of the germline genotype which is selected against, is then $1-s$. Thus, after one generation, the new mutation frequency, $M_{q_{t}}$, in terms of the recessive germline allele $q_{t}$ is given by equation \ref{eq:basic_mut_frequency}, which is,

\begin{equation}
      M_{q_{t}} = \frac{q_{t}-sq_{t}^2}{1-sq_{t}^2}.
\end{equation}

\noindent Hence, change in mutation frequency, $\Delta M_{q_{t}}$, resulting in a small time interval $\Delta t$ of selection is,

$$\Delta M_{q_{t}} = [M_{q_{t}}-q_{t}] \Delta t = -\frac{sq_{t}^2(1-q_{t})}{1-sq_{t}^2} \Delta t.$$ 

\noindent Let $T_{0}$ and $T$ be the time of tumor initiation and the time of tumor biopsy, respectively. Then,

$$M(q_{T}) = \int_{T_{0}}^{T}dM_{q_{t}} = -\int_{T_{0}}^{T}\frac{sq_{t}^2(1-q_{t})}{1-sq_{t}^2}dt.$$

\noindent Denoting $u=q_{t}$, and applying inverse function theorem,

$$\frac{du}{dt}=q'_{t}=\frac{1}{[q^{-1}]'(u)}=\frac{1}{F'(u)}$$

\noindent where $q(q_{t_i})=q_{t_i}$ for any time point $t_i$ and $F=q^{-1}$. We see that, if $u_{0} = q_{T_0}$ and $u_{T} = q_{T}$ are initial and final frequencies, then

\begin{equation}
    \label{eq:main_equation}
    M(u_{T}) = -\int_{u_{0}}^{u_{T}}\frac{su^2(1-u)}{1-su^2}F'(u)du
\end{equation}

\noindent This formula allows us to express the number of mutations purely in terms of allele frequencies. If $\lambda$ is the fraction of reduction of germline alleles due to the selection pressure $s$, we expect $F$ to follow the function,

$$F(u)=\frac{1-\lambda u}{u-\lambda u} \Rightarrow F'(u) = -\frac{1}{(1-\lambda)u^2}$$

\noindent and therefore, equation \ref{eq:main_equation} is given by,

$$M(u_{T}) = \frac{s}{1-\lambda}\int_{u_{0}}^{u_{T}}\frac{1-u}{1-su^2}du \approx s_{\lambda}\int_{u_{0}}^{u_{T}}(1-u)du = s_{\lambda}\left(u_{T}-\frac{u_{T}^2}{2}\right)+s_{\lambda}C$$

\noindent where $s_{\lambda}=s/1-\lambda$ can be interpreted as \textit{mean effective selection coefficient} favoring mutant genotypes. If the coefficient is extremely small at $T_{0}$, then $s_{\lambda}C \approx 0$, and hence the number of mutations in terms of the observed germline allele frequency $q$ at time $T$ is given by,

\begin{equation}
    \label{eq:M_q_equation}
    M(q)\approx s_{\lambda}\left(q-\frac{q^2}{2}\right)
\end{equation}
\noindent and $s_{\lambda}$ is therefore,

\begin{equation}
    \label{eq:s_lambda_equation}
    s_{\lambda} \approx M(q)\left[q-\frac{q^2}{2}\right]^{-1}
\end{equation}

Also, from equation \ref{eq:basic_mut_frequency}, since we have,

\begin{equation}
    \label{eq:M_p_equation}
     M(p) \approx p+s_{\lambda}p(1-p)^2, 
\end{equation}

\noindent if $s_{\lambda}$ is very small or when $p$ is small, the number of mutations $M(p)$ in terms of the mutant allele frequency $p$ can be approximated by $s_{\lambda}p+p$. Therefore, we expect $s_{\lambda}$ to be correlated with the distribution of the mutant allele frequencies for small values of $s_{\lambda}$. However, we note from equation~\ref{eq:s_lambda_equation} and equation~\ref{eq:M_p_equation} that extremely small values of $q$ or $M(p)=0$ will make $s_\lambda$ unbounded or negative, and hence correlation may not be valid.\\

\textbf{3. Results}

\begin{wrapfigure}{R}{0.50\textwidth} 
    \centering
    \captionsetup{justification=centering}
    \includegraphics[width=0.50\textwidth]{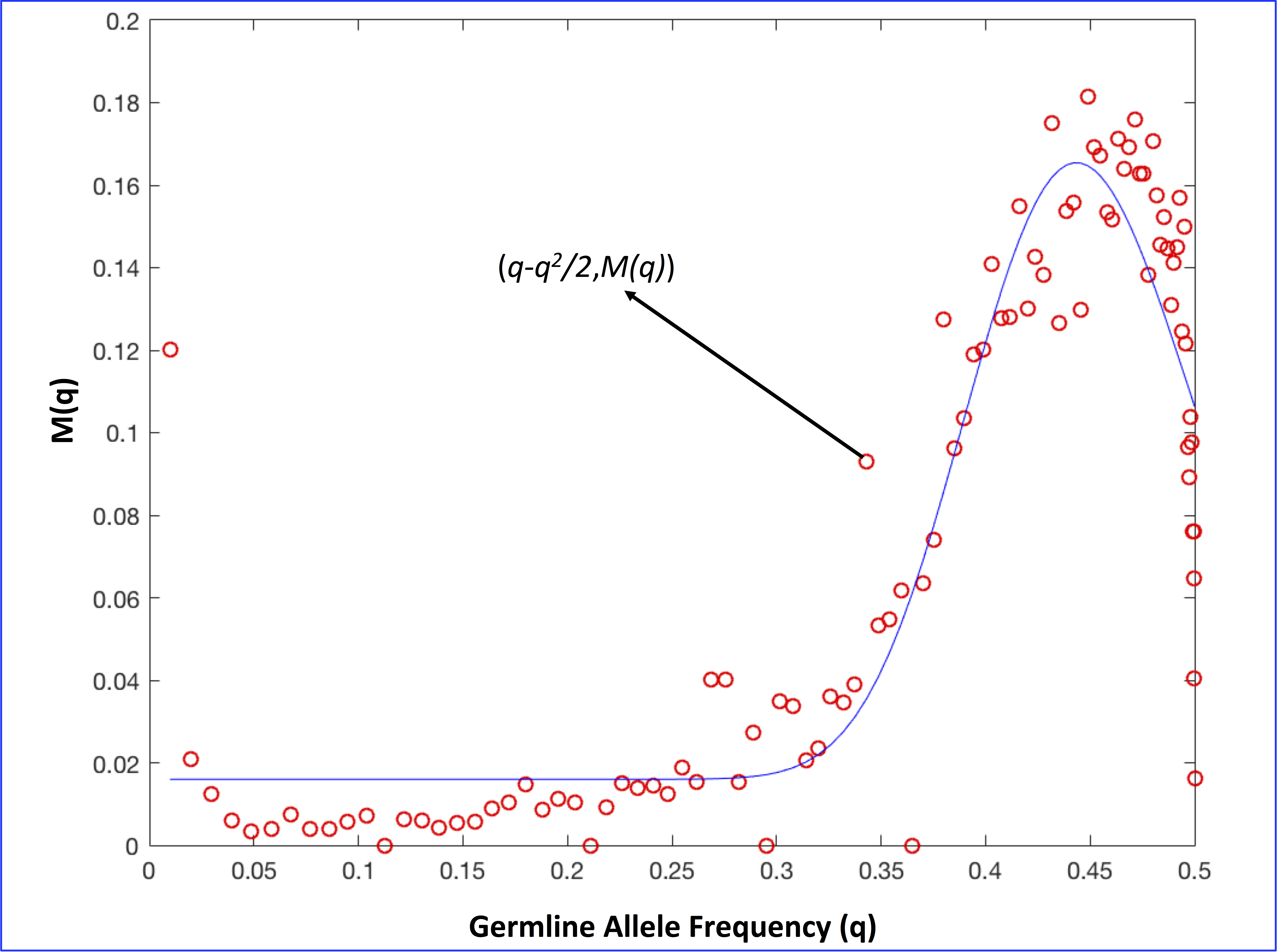}
    \caption[]{Scatter plot of $(q-q^2/2,M(q))$ and the gamma distribution curve fit}
    \label{fig:gamma_fit}
\end{wrapfigure}

We determined the selection coefficents $s_{\lambda}$ for mutations in oncogenes to see if they are correlated with the distribution of the mutant allele frequencies. To do this, we first identified $574$ proto-oncogenes from the Uniprot database~\cite{uniprot} out of which $236$ were exclusive to {\textit{Homo sapiens}}. A total of $42,525$ mutations in these genes were queried from the cBio portal~\cite{cBio1,cBio2} and $25,848$ mutations for which mutant allele frequencies were available, was retained (Supplementary Table 1 available upon request). The germline allele frequencies were computed by subtracting the mutant allele frequencies from $1$ and $M(q)$ was normalized with its standard Eucledian norm.

Equation~\ref{eq:M_q_equation} essentially states $s_{\lambda}$ should be correlated with the random variable $M(q)$ under the random variable $q-q^2/2$. Therefore, it is natural to fit the data $(q-q^2/2,M(q))$. Since two fitness distributions, the gamma, and the exponential have been traditionally applied to model selection coefficients~\cite{gillespie}, we employed both functions. Consistent with the fitness function of deleterious mutations in human polymorphisms~\cite{eyre}, the gamma distribution curve fitted well with minimal residual error of $6\times10^{-4}$ with fitting function with respect to $x=q-q^2/2$, given by,
\begin{equation*}
    g(\alpha,\beta,\rho,\delta;x) = \frac{\rho}{\beta^{\alpha}\Gamma(\alpha)}x^{\alpha-1}e^{\frac{-x}{\beta}} + \delta
\end{equation*}

where shape, scale, amplitude and offset parameters, $\alpha$, $\beta$, $\rho$ and $\delta$ were identified as $68.17$, $0.006$, $0.02$, $0.01$, respectively. MATLAB's {\tt{nlinfit}} was used to fit the data with initial conditions $[1,1,0.5,1]$ for the parameters.

\begin{figure}[h!] 
    \centering
    \includegraphics[width=0.80\textwidth]{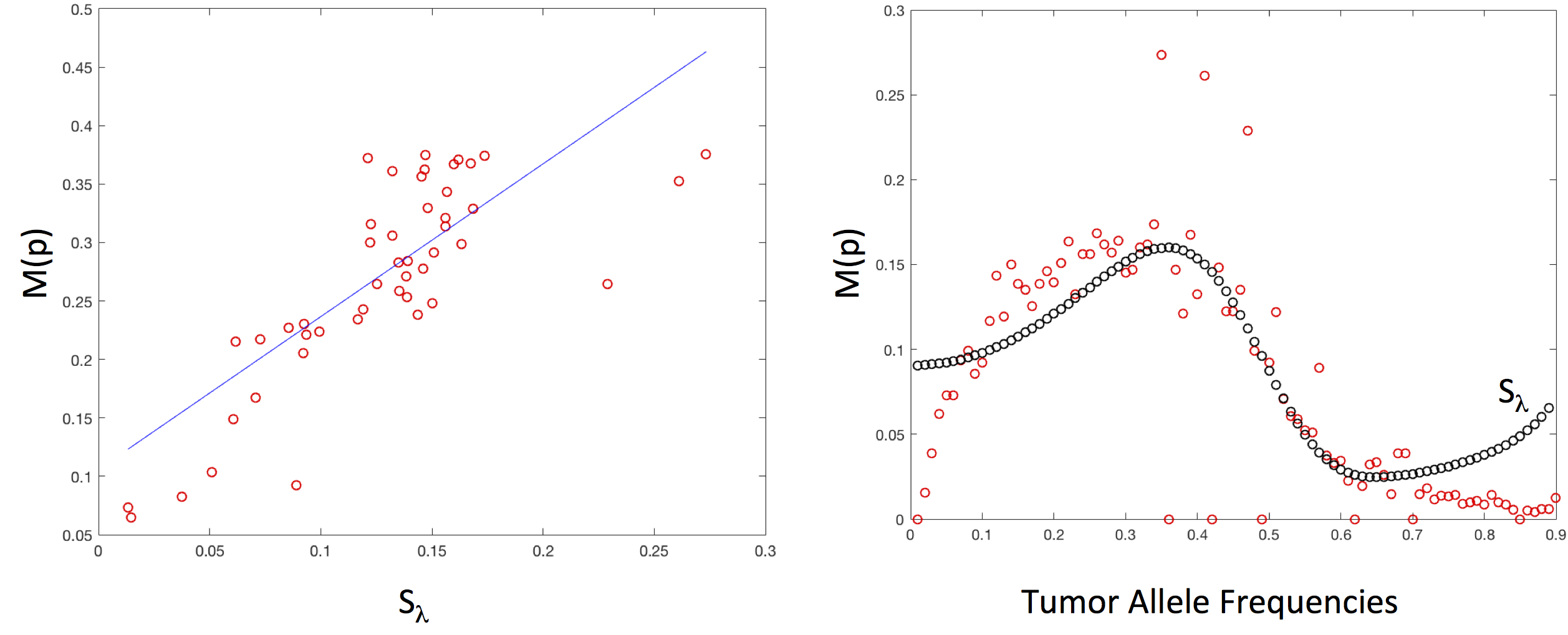}
    \caption{Correlation between $s_{\lambda}$ and $M(p)$}
    \label{fig:correlation}
\end{figure}

Figure~\ref{fig:gamma_fit} shows the scatter plot of $(q-q^2/2,M(q))$ and the gamma distribution curve fit. This function allows us to compute $s_{\lambda}$ for each mutation through equation~\ref{eq:s_lambda_equation}, i.e., $s_{\lambda}=g(\alpha,\beta,\rho,\delta;x)/x$, where $x$ is the transformed allele frequency, $q-q^2/2$. Since $s_{\lambda}$ coefficients are as much as the number of mutations and $M(p)$ is restricted by the frequency bin size, to find the correlation, we sub-sampled $s_{\lambda}$ for $10,000$ iterations and considered mean correlation. Also, for the reasons discussed following equation~\ref{eq:M_p_equation}, mutant allele frequencies greater than $0.90$ and $M(p)=0$ were not considered. The mean correlation was determined to be $0.79$ with mean p-value $2.3e^{-12}$. We also determined the line of best fit (MATLAB's {\tt{polyfit}} routine with degree $1$) to infer the slope to find the optimal $s_{\lambda}$ that would fit with the mutant allele frequencies. Figure~\ref{fig:correlation} (a) shows the correlation between $s_{\lambda}$ coefficients and $M(p)$ and figure~\ref{fig:correlation} (b) shows optimal $s_{\lambda}$ that fits the mutant allele frequencies. 

\textbf{4. Discussion}

Good correlation between selection coefficients $s_{\lambda}$ and the mutant allele frequencies explain why the frequencies are centered around $0.40$ and rapidly decrease. Selection coefficients are maximized around this region and reduce exponentially.

In population genetics it is known that at low frequencies under dominance, selection for/against dominant alleles follow $O(p)$ and selection for/against recessive alleles grow at $O(q^2)$. This is because equation~\ref{eq:basic_mut_frequency} tells us that,
\begin{equation*}
\Delta f(p,q) = \pm \frac{spq^2}{1-sq^2}.
\end{equation*}

While this is true when the unit of measure is generation time, integrating the selection equation with respect to time establishes that at all frequencies selection for the dominant alleles with respect to change in recessive allele frequency is linear. This can be seen by differentiating equation~\ref{eq:M_q_equation}:
\begin{equation}
    \label{eq:dmq}
    \frac{dM(q)}{dq} = s_{\lambda}(1-q) = s_{\lambda}p.
\end{equation}

Similar analysis on selection against recessive allele with respect to dominant allele frequency would reveal (See Supplementary Method $1$),
\begin{equation}
    \label{eq:dmp}
    \frac{dM(p)}{dp} = -s_{\theta}\frac{q^2}{p}.
\end{equation}

where $-s_{\theta} = -s(1-\lambda)$ can be interpreted as effective selection against recessive alleles. These formulas, equations~\ref{eq:dmq} and~\ref{eq:dmp}, as relative increase and decrease, define natural selection under complete dominance. Equation~\ref{eq:dmq} is natural after all - for a given dominant allele frequency, the rate of change of dominant alleles in terms of (the loss of) germline alleles, should be proportional to the amount of selection that takes place. While both equations are consistent with known observation under low frequencies, equation~\ref{eq:dmp} suggests that at low frequencies of the dominant alleles, selection against the recessive alleles act in a power-law like manner, demonstrating why positive (natural) selection is a potent force. It is worth noting that this power-law growth of relative fitness has been observed in long term evolution experiments~\cite{lenski1,lenski2}. Under complete dominance, selection would maximize the dominant alleles at all costs. 

Equations~\ref{eq:dmq} and~\ref{eq:dmp} also allows us to compute the rate of change of gene/mutation frequencies with respect to time. If $\alpha$ and $\beta$ are the rate of growth of dominant and recessive alleles, respectively, we can expect an exponential growth. Therefore,

$$\frac{dp}{dt} = \alpha p;  \hspace{3mm} \frac{dq}{dt} = \beta q.$$

Hence, denoting $M(q)$ by $M_q$ and $M(p)$ by $M_p$, we see,

\begin{equation}
    \label{eq:dmq_t}
    \frac{dM_q}{dt} = \frac{dM_q}{dq}\cdot \frac{dq}{dt} = s_{\lambda}\beta pq,
\end{equation}

and

\begin{equation}
    \label{eq:dmp_t}
    \frac{dM_p}{dt} = \frac{dM_p}{dp}\cdot \frac{dp}{dt} = -s_{\theta}\alpha q^2.
\end{equation}

Thus, if the genes/mutations don't directly depend on time, the total derivative, i.e., selection acting over time, is given by, 

\begin{equation*}
    \label{eq:dmpq_t}
    \frac{dM}{dt} = s_{\lambda}\beta pq -s_{\theta}\alpha q^2.
\end{equation*}

Further, equations~\ref{eq:dmq_t} and~\ref{eq:dmp_t}, helps us write the general equation for evolution through natural selection for diploid genomes under complete dominance,

$$ \alpha^2 s_{\theta} dq dM_q + \beta^2 s_{\lambda} dp dM_p = 0. $$

In the context of cancer, especially in the evolution of mutant clones in oncogenes when mutant allele frequencies are small, the normal linear growth of the mutant alleles along with the power-law like loss of recessive germline alleles will doubly accelerate the progression of the mutant clones, possibly contributing to heterogeneity in the presence of competing mutations. Similarly, when recessive germline allele frequencies are small, and they grow quadratically, the progression of the mutant alleles will proceed linearly, still dominating and possibly conferring more resistance to therapy and giving rise to metastatic clones. Therefore, incorporating selection measures in evaluating functional impact of the mutations and assessing the aggressiveness of the tumors by taking the degree of selection into account will lead to better understanding of this complex evolutionary disease.

\textbf{Acknowledgements}

The author likes to acknowledge Dr. Friedrich Philipp who suggested the use of inverse function theorem in an online forum.

\textbf{Notes}

The MATLAB routine used to generate the results is available upon request.

\printbibliography

\vspace{0.3cm}
\textbf{Copyrights}

Copyright for this article is retained by the author(s), with first publication rights granted to the journal.

This is an open-access article distributed under the terms and conditions of the Creative Commons Attribution license (http://creativecommons.org/licenses/by/3.0/).

\newpage
\begin{center}
{\textbf{Supplementary Figure 1}}\\[20pt] 
\end{center}

\begin{figure}[h!] 
    \centering
    \includegraphics[width=0.80\textwidth]{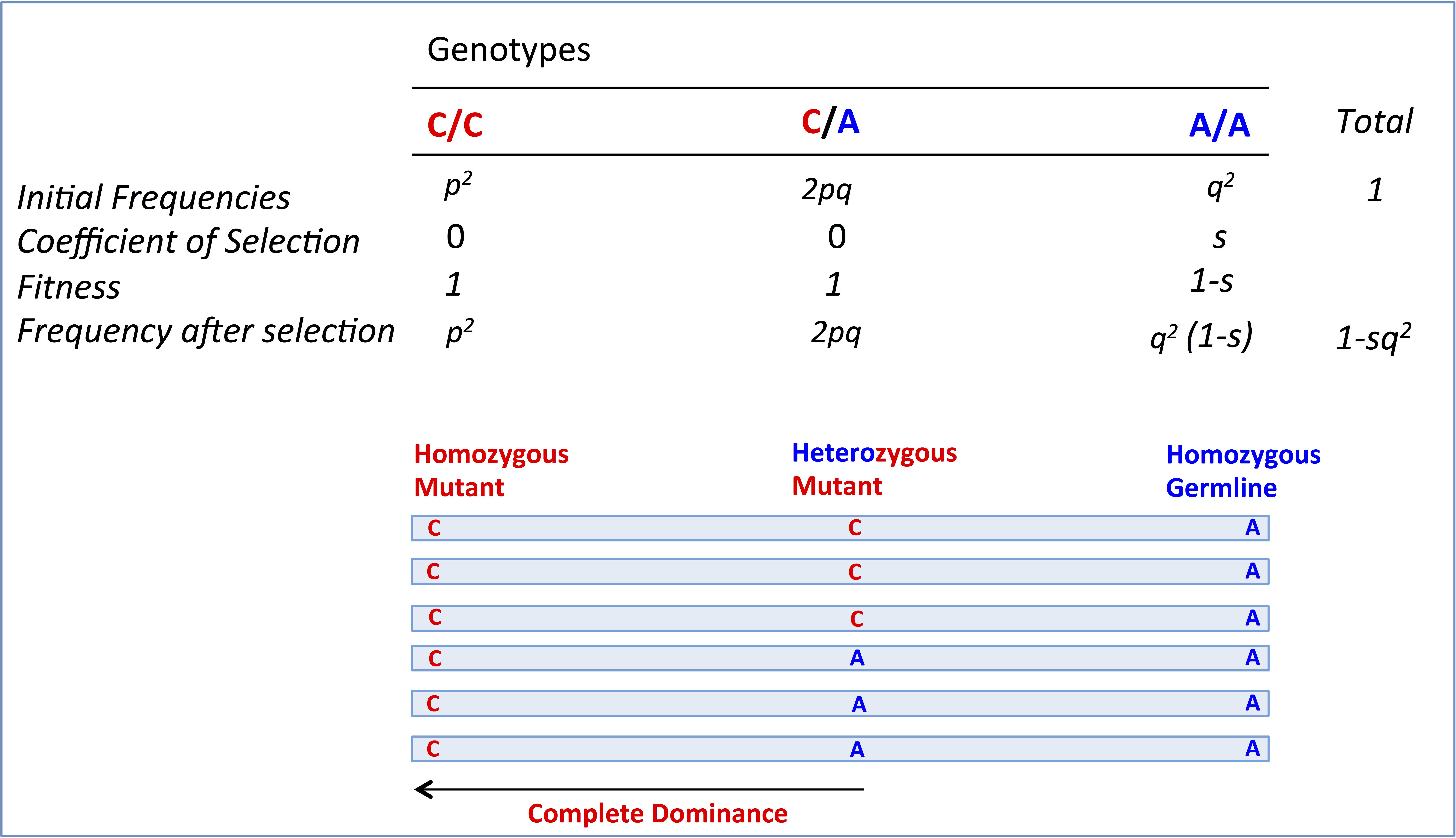}
    \caption{Selection under complete dominance}
    \label{fig:dominance}
\end{figure}

\newpage

\begin{center}
{\textbf{Supplementary Method 1}}\\[20pt] 
\end{center}

In the case of dominant allele frequencies,

\begin{equation*}
      M_{p_{t}} = \frac{p_{t}}{1-sq_{t}^2}.
\end{equation*}

\noindent Hence, change in mutation frequency, $dM_{p_t}$, resulting in a small time interval $dt$ of selection is,

\begin{equation}
    \label{eq:dmpt}
    dM_{p_{t}} = (M_{p_{t}}-p_{t}) dt = +\frac{sp_{t}q_{t}^2}{1-sq_{t}^2} dt.
\end{equation}

\noindent Denoting $v=p_{t}$, and applying inverse function theorem,

$$\frac{dv}{dt}=p'_{t}=\frac{1}{[p^{-1}]'(v)}=\frac{1}{G'(v)}$$

\noindent where $p(p_{t_i})=p_{t_i}$ for any time point $t_i$ and $G=p^{-1}$. 

\noindent If $\lambda$ is the fraction of reduction in the germline alleles due to the selection pressure $s$, we expect $G$ to follow the function,

$$G(v)=\frac{1-\lambda (1-v)}{v} \Rightarrow G'(v) = -\frac{1-\lambda}{v^2}$$

\noindent and therefore, equation \ref{eq:dmpt} is given by,

\begin{equation*}
    \frac{dM(p)}{dp} = -s(1-\lambda)\frac{q^2}{p[1-s(1-p)^2]} \approx -s(1-\lambda)\frac{q^2}{p}
\end{equation*}

\end{document}